\documentclass[fleqn,10pt]{wlscirep}
\usepackage{graphicx,amsmath,amssymb,verbatim,color}
\newcommand{\be}{\begin{equation}}
\newcommand{\ee}{\end{equation}}

\title{Layer-switching cost and optimality in information spreading on multiplex networks}
\author[1,2,*]{Byungjoon Min}
\author[1,*]{Sang-Hwan Gwak}
\author[1]{Nanoom Lee}
\author[1,$\dagger$]{K.-I.~Goh}
\affil[1]{Department of Physics, Korea University, Seoul 02841, Korea}
\affil[2]{Levich Institute and Physics Department, City College of New York, New York, NY 10031, USA}

\affil[*]{these authors contributed equally to this work}
\affil[$\dagger$]{corresponding author: kgoh@korea.ac.kr}

\begin{abstract}
We study a model of information spreading on multiplex networks, in which agents interact through multiple interaction channels (layers), say online vs.\ offline communication layers, subject to layer-switching cost for transmissions across different interaction layers.  The model is characterized by the layer-wise path-dependent transmissibility over a contact, that is dynamically determined dependently on both incoming and outgoing transmission layers. We formulate an analytical framework to deal with such path-dependent transmissibility and demonstrate the nontrivial interplay between the multiplexity and spreading dynamics, including optimality. It is shown that the epidemic threshold and prevalence respond to the layer-switching cost non-monotonically and that the optimal conditions can change in abrupt non-analytic ways, depending also on the densities of network layers and the type of seed infections. Our results elucidate the essential role of multiplexity that its explicit consideration should be crucial for realistic modeling and prediction of spreading phenomena on multiplex social networks in an era of ever-diversifying social interaction layers.
\end{abstract}

\begin{document}
\flushbottom
\maketitle
\thispagestyle{empty}

\section*{Introduction}
Networks are penetrating ever more deeply through every facet of individual lives and societal functions \cite{chiang}. 
At its center, the explosive rise of social media driven by the information communication
technology or ICT revolution has profoundly transformed the landscape of human interactions.
Human interactions mediated by social media could defy the spatial and temporal limitations of traditional communications in an unprecedented  way, thereby offering a qualitatively new layer of social interaction, which coexists and cooperates with existing interaction layers to redefine the multiplex social networks \cite{rumor_higgs,twitter,dubai}. 
In addition, networks with different types of edges categorized by their relationships have been studied for a long time in social network analysis \cite{verbrugge,padgett,szell}. 
These multiple interaction channels or network layers in a multiplex system do not function completely
autonomously nor dependently; while each layer can support some function within its scope, 
it is the crosstalk and interplay between these layers that  can fulfill the full functionality of the system and could give rise to
nontrivial and unanticipated collective outcomes such as the recently uprising civil movements in the Middle East \cite{dubai}.
This poses theoretical challenge as well to extend existing single-network framework \cite{network,cohen-havlin,dynamics} 
by formulating and disseminating the role of multiplex layers that do not always play independent roles in network structure and dynamics, the understanding of which is beginning to be culminated~\cite{NoN,multilayer_jcn,multilayer_physrep,epjb}.

Epidemic processes on networks are one of the most actively developed branches in complex network theory~\cite{epidemic_rmp},
which can address not only the spreading of infectious diseases 
but also many other contagious phenomena such as information and rumor spreading on social networks. 
A few recent studies on epidemic spreading beyond the single-network framework have been performed under various terms like overlay networks \cite{funk}, multitype networks~\cite{allard}, interconnected networks~\cite{dickison,mendiola}, interdependent networks \cite{son}, interacting networks \cite{sanz}, and multiplex networks \cite{arenas,moreno}. 
Cascade processes have also been studied on multilayer, interdependent, and multiplex networks \cite{buldyrev,gao_nphys,tan-cascading,brummitt,KM_2014,viability,reis}.
(For details of these terms and their similarity and differences, the reader is referred to the comprehensive table compiled in Ref.~\citen{multilayer_jcn}.) 
Here we study an epidemic-based information spreading model framework on multiplex social networks, distinguished from existing models by the presence of the layer-switching cost, describing the overburden or surcharge for transmissions that proceed by crossing different layers compared to those proceeding as confined within the same layer.
For example, when one received new information through an online social medium, say Twitter, 
she would more likely spread it again through Twitter as it is most handy, than would do it over 
other online media, such as e-mail, let alone over an offline social network, as it would require additional effort and/or accompany spatiotemporal delay in switching the medium (network layer).
Indeed, early studies using data from Twitter and weblogs have shown that the information diffusion structure is highly platform-dependent \cite{jkwon,jyang}.
Despite being commonplace, the layer-switching cost has not yet been explicitly addressed in multiplex spreading dynamics and thus its implication is not elucidated systematically.

In this paper, we show that this commonplace factor of layer-switching cost 
can significantly and nontrivially modify information spreading dynamics on multiplex social networks.
Most fundamentally, it introduces the path-dependent transmissibility over a contact 
that is dynamically determined depending on both incoming and outgoing transmission layers, 
which requires a new theoretical formalism beyond the standard ones~\cite{newman,miller,kenah}.
We formulate a generating-function based theory to cope with such path-dependent transmissibility in locally-treelike networks. 
Using both analytical calculations and numerical simulations, various consequences of the layer-switching cost, and thus the path-dependent transmissibility, are revealed. These include
the existence of trade-off  between the infection rates along the same layer
and across difference layers to optimize information spreading for a given average infection rate over different channels 
and the nearly-confined spreading  within the dominant layer when the layer-switching cost is large enough.
Our study elucidates how the network multiplexity and the layer-switching cost can alter the information spreading
dynamics in non-trivial way and thereby suggests that the modeling neglecting the multiplex social interactions into an aggregated one could potentially mislead to inaccurate conclusions.

\begin{figure}
\includegraphics[width=15.5cm]{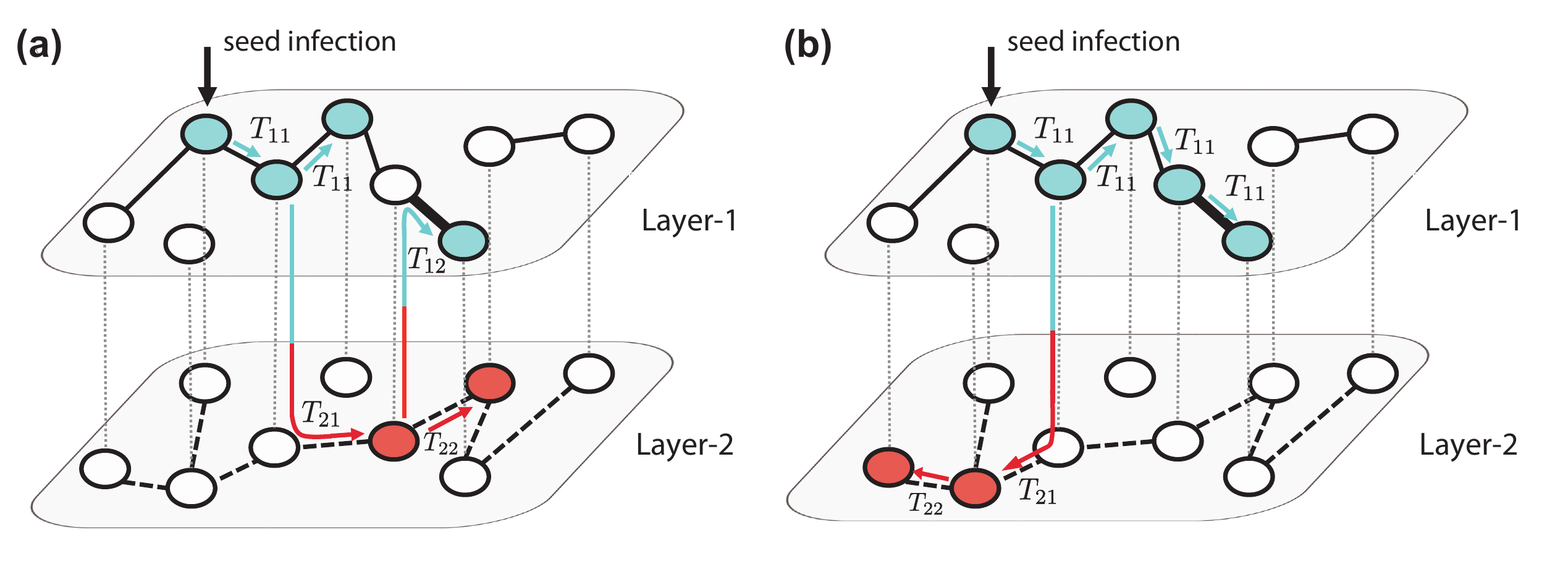}
\caption{
Schematic illustration of the information spreading model with layer-switching cost on the multiplex network with two layers. In this example, the spreading starts with the seed infection in layer 1. Subsequent spreading proceeds with the transmissibility $T_{ji}$ as the transmission in layer $j$ is preceded by the transmission in layer $i$. Two possible sample spreading trajectories are shown for illustration in (a) and (b), respectively. Nodes are colored according to the type of transmission through which they are infected (blue for type-1 and red for type-2 transmission). Due to the path-dependency, the transmissibility of a given link is not fixed {\it a priori} but can take different value, as exemplified by the thick link in this figure. 
}
\label{fig:our_model}
\end{figure}

\section*{The Model }

We take account of the effect of network multiplexity and layer-switching cost by introducing the layer-wise path-dependent infection rates. 
Given more than one layer through which the information or disease spreads, the infection rate for a link (contact) in one layer depends not only on the current layer but also on which layer the information or disease has transmitted from (as in the case of the Twitter example in the previous section). 
To implement this idea specifically, we construct a model based on the prototypical susceptible-infected-recovered (SIR) model framework for epidemic spreading \cite{anderson} taking place on multiplex networks with more than one layer.
In the SIR model, each node is in one of three states, susceptible, infected, or recovered (or removed).
An infected node can spread information (or rumor or disease) to a susceptible neighbor with the infection rate $\beta$,
and each infected node is recovered after a time $\tau$ from the moment of infection.
We here assume that the probability distribution of the recovery time $\tau$ is sharply peaked,
and so well-described by the delta function.
The probability that an infected agent infects its neighbors before recovery, denoted by $T$, 
is called the transmissibility, which is, under the above assumption, given by $T=1-e^{-\beta \tau}=1-e^{-\lambda}$ \cite{newman}, defining the dimensionless parameter $\lambda\equiv\beta\tau$ as the effective infection rate.
The average fraction of recovered nodes in the stationary state ($t\to\infty$ limit), $\rho$, is called the prevalence and is the main observable in the spreading process.

For the spreading process on a multiplex network with in general $\ell$ layers, 
we define the type-$i$ transmission to be the infection event in which the infection occurs through a link in layer $i$ ($i=1,2,\dots,\ell)$. 
The key feature of our model is that the infection rate over a link depends on the types of both incoming (preceding) and outgoing (current) infection layers.
To be specific, when a node is infected via a type-$i$ transmission, 
then the effective infection rate for the infection through the same layer link (type-$i$ transmission) is $\lambda_{ii}$ whereas that through a link in different layer $j$ (type-$j$ transmission) is $\lambda_{ji}$, where these infection rates are different in general. With the interest on the effect of layer-switching cost, we mainly consider the case $\lambda_{ii}\ge\lambda_{ji}$, yet the model framework itself does not impose any such constraint.
As a consequence of layer-wise path-dependent infection rates in our model, the transmissibilities $T_{ij}$ become accordingly dependent on the types of both incoming and outgoing infection layers. Therefore, the transmissibility through a given link is not fixed {\it a priori} but can change as the spreading dynamics proceeds, as exemplified in Fig.~1.

\section*{Analytical framework}

In this section we develop an analytical framework for our model applicable to the case of the multiplex network of locally-treelike random network layers,  based jointly on the well-established single-network framework for SIR models~\cite{newman,miller,kenah} as well as the percolation on multiplex networks \cite{leicht,quark,quark_NoN,min-robustness,feng}.
(While our theory is strictly derived from the assumption of locally-treelike random networks, the theoretical approach may also apply to some non-treelike real networks as reported in Ref.~\citen{melnik}.)
For the sake of explicit illustration, our discussion proceeds for the simplest case of 2-layer (duplex) networks, and the generalization to $\ell>2$ layers is straightforward.

\subsection*{Outbreak size}
We first consider the average epidemic size once the epidemic 
outbreak occurs, denoted $S$ and called the outbreak size,
equivalent to the average nonzero final fraction of recovered nodes.
According to the standard theory \cite{miller,kenah}, in order to estimate the outbreak size $S$ one needs the incoming transmissibility of each type of links, which however is not given {\it a priori} in our model.
In our model, the incoming transmissibility for a link is not assigned inherently and definitively
but determined dynamically and dependently on the transmission channel by which the infecting node had become infected.
In what follows we tackle this difficulty by using a method based on the self-consistency argument to infer the effective incoming transmissibility yielding the same outbreak size $S$ as the original problem.

In order to infer the effective incoming transmissbility for each kind of links, we first estimate the probability $\pi_{ij}$ that an infected node reached by following a randomly-chosen type-$i$ link had been infected by a type-$j$ transmission.
This probability $\pi_{ij}$ can be expressed in terms of the probability $\pi_{ij}^{(k_1,k_2)}$ that an infected node with the multiplex degree $(k_1,k_2)$ reached by following a randomly-chosen type-$i$ link had been infected by a type-$j$ transmission as 
$\pi_{ij}=\sum_{k_1=0,k_2=0}^{\infty} \frac{k_i p(k_1,k_2)}{z_i} \pi_{ij}^{(k_1,k_2)}$, where $z_i$ is the mean degree of the layer $i$.
In our model, there are two different possible ways that a node is infected by a type-$i$ transmission:
the node could be infected by a neighbor which had been infected either by a type-$i$ or type-$j$ transmission, with respective transmissibilities, $T_{ii}$ and $T_{ij}$. 
For locally-treelike layers this consideration leads that  
$\pi_{ij}^{(k_1,k_2)}$ and $\pi_{ii}^{(k_1,k_2)}$ are respectively proportional to 
$k_j (T_{ji} \pi_{ji} +T_{jj} \pi_{jj})$ and $(k_i-1) (T_{ii} \pi_{ii} +T_{ij} \pi_{ij})$.
Summing up for the multiplex degree and properly normalizing lead to the coupled self-consistency equations for $\pi_{ij}$'s, given by 
\begin{align}
\pi_{ii}&=\frac{\kappa_i}{\pi_i}(T_{ii} \pi_{ii}+T_{ij}\pi_{ij}),\nonumber \\
\pi_{ij}&=\frac{\mathcal{K}_i}{\pi_i}(T_{ji}\pi_{ji}+T_{jj}\pi_{jj}),
\end{align}
where 
\begin{equation*}
\kappa_i=(\langle k_i^2\rangle-z_i)/z_i, \quad \mathcal{K}_i=\langle k_i k_j\rangle/z_i, 
\end{equation*}
and 
$\pi_{i}$ is the normalization factor imposed by $\pi_{ii}+\pi_{ij}=1$, for distinct $i,j\in\{1,2\}$.
Solving these equations for $\pi_{ij}$'s with given $p(k_1,k_2)$ and $T_{ij}$'s (equivalently, $\lambda_{ij}$'s), one can obtain the effective average incoming transmissibility through 
the type-$i$ link, denoted $\tilde{T_i}$ and given by
\begin{eqnarray} 
\tilde{T_i}=T_{ii}\pi_{ii}+T_{ij}\pi_{ij}.
\end{eqnarray} 

What is achieved thus far is to transform the original model into an equivalent (with respect to $S$) SIR model with (path-independent) transmissibilty $\tilde{T_i}$ in each layer $i$.
The outbreak size of the transformed model can be found in the standard way, by adapting the methods developed for percolation in multiplex networks \cite{leicht,quark,quark_NoN,min-robustness}.
Let $G_0(x,y)$ be the generating function of the joint degree distribution 
$p(k_1,k_2)$, $G_0(x,y)=\sum_{k_1=0,k_2=0}^\infty p(k_1,k_2) x^{k_1}y^{k_2}$. 
The generating function $G_0(x,y;p,q)$ of the joint distribution of the
numbers of occupied edges when the edges are independently occupied with 
the probability $p$ in layer 1 and $q$ in layer 2 can be written as $G_0(x,y;p,q)=G_0\left(1+(x-1)p,1+(y-1)q\right)$. 
In the transformed SIR model the edges in layer $i$ are occupied with probability $\tilde{T}_i$, so the probability that a node reached by following a randomly-chosen type-$i$ link 
does not belong to the epidemic outbreak, denoted $x_i$, is given by \cite{newman}
\begin{eqnarray}
x_{i}=\frac{1}{z_i \tilde{T_i}}\frac{\partial}{\partial x_i}G_0(x_1,x_2;\tilde{T}_1,\tilde{T}_2)
\end{eqnarray}
with $i=1,2$.
Finally, the outbreak size $S$ (that is, the probability that a random-chosen infected node belongs to the giant connected component of infected nodes) can be obtained as
\begin{eqnarray}
S=1-G_0(x_1,x_2;\tilde{T}_1,\tilde{T}_{2}),
\end{eqnarray} 
with $x_i$'s being the physical solution of Eq.~(3).

\subsection*{Outbreak probability}
The outbreak probability can be in general different with 
the outbreak size due to the effective directionality induced by the path-dependent transmissibility in our model~\cite{kenah,miller}.
In order to obtain the outbreak probability, we can follow the path-dependent transmissibilities determined by the incoming and the outgoing transmission channels explicitly. 
Let $y_i$ be the probability that a node infected by type-$i$ transmission 
does not lead to an epidemic outbreak. 
Similarly to Eq.~(3), $y_i$'s satisfy the coupled self-consistency equations 
\begin{align}
y_{i}&=\frac{1}{z_i T_{ii}}\frac{\partial}{\partial y_i}G_0(y_1,y_2;T_{1i},T_{2i}),
\end{align} with  $i=1,2$. 
The probability $P_i$ that a type-$i$ seed infection gives rise to an epidemic outbreak (that is, the infection spreads indefinitely) is then given by
\be
P_i=1-G_0(y_{1},y_2;T_{1i},T_{2i}) ,
\ee
with $y_i$'s being the physical solution of Eq.~(5).
Note that unlike the outbreak size $S$, the outbreak probability $P_i$ depends on which layer the epidemic is initiated from.

\subsection*{Epidemic threshold}
The epidemic threshold can be obtained by 
the linear stability criterion of the trivial fixed point $(y_1,y_2)=(1,1)$ of Eq.~(5).
The condition of the epidemic outbreak requires the largest eigenvalue $\Lambda$ of 
the Jacobian matrix $\mathbf{J}$ of Eq.~(5) at $(1,1)$ to be larger than unity, $\Lambda>1$,
which is the condition of the fixed point being unstable.
The Jacobian matrix $\mathbf{J}$ at $(1,1)$ can be simply expressed as 
\begin{eqnarray}
\left( \begin{array}{cc}
T_{11} \kappa_1 & T_{21} \mathcal{K}_1 \\
T_{12} \mathcal{K}_2 & T_{22} \kappa_2 
\end{array} \right).
\end{eqnarray}
The largest eigenvalue $\Lambda$ can be explicitly calculated as
\begin{align}
\Lambda= \frac{1}{2}\bigg[ T_{11} &\kappa_1 +T_{22}\kappa_2 +\sqrt{(T_{11}\kappa_1-T_{22}\kappa_2)^2+4 T_{12} T_{21} \mathcal{K}_1 \mathcal{K}_2}\bigg]~.
\end{align}
Note that $\Lambda\ge \max(T_{11}\kappa_1,T_{22}\kappa_2)$, meaning that the epidemic
threshold of the multiplex network cannot be larger than the epidemic thresholds of individual layers.

\subsection*{Comparison with numerical simulations}

To verify the validity of the proposed analytical framework, we compare the analytical calculation with the numerical simulation results. The numerical simulation of our model proceeds as follows.
Initially, all nodes are susceptible except for one randomly-chosen seed
which is assumed to be infected by a type-$i$ transmission (that is, infected through layer $i$).
Infected nodes transmit the disease to their susceptible neighbors with the infection rates $\lambda_{ji}$
determined by both the incoming channel $i$ and the outgoing channel $j$. 
Each infected node recovers after a fixed recovery period, $\tau$.
The spreading process proceeds until all infected nodes in the network recover, which completes one independent run of the numerical simulation.
After many independent runs, we compute the outbreak probability $P_i$ as the fraction of runs ending up with the fraction of recovered nodes above the prescribed threshold (chosen to be $1\%$ in our numerical simulation).
Likewise the outbreak size $S$ is computed as the average of the fraction of recovered nodes above the threshold.
The prevalence due to type-$i$ seed infection, $\rho_i$, is computed as $\rho_i=P_i S$.

\section*{Duplex ER networks}

To gain further insights on the role of layer-switching cost, we elaborate further on analyzing the model on the 2-layer network formed by two independently-constructed Erd\H{o}s-R\'enyi (ER) layers (henceforth the duplex ER network, for short), for which one can obtain the analytical results in an explicit form. 
Mean degrees of two ER layers are denoted $z_1$ and $z_2$, respectively. 
To focus on the effect of layer-switching cost, we further simplify the parameter setting such that the infection rates within the same layer are equal as $\lambda_{11}=\lambda_{22}\equiv\lambda_s$ (``s" for same) and  similarly for the infection rates across different layers as $\lambda_{12}=\lambda_{21}\equiv\lambda_d$ (``d" for different).
Corresponding transmissibilities are given by $T_{i}=1-e^{-\lambda_{i}}$, where $i$ is either $s$ or $d$.
We further assume $\lambda_{s}\ge \lambda_{d}$, so that the information spreading along the same layer is easier than that across layers, in compliance with the concept of layer-switching cost.

When the two layers are randomly-coupled, the effective incoming transmissilibities can be calculated under this simplified setting as 
\begin{align}
\tilde{T}_{1}=\frac{T_s}{2}-\frac{z_2 T_d}{z_1-z_2} + \sqrt{\frac{T_s^2}{4}+\frac{z_1 z_2 T_d^2}{(z_1-z_2)^2}} \quad\textrm{and}\quad
\tilde{T}_{2}=\frac{T_s}{2}+\frac{z_2 T_d}{z_1-z_2} - \sqrt{\frac{T_s^2}{4}+\frac{z_1 z_2 T_d^2}{(z_1-z_2)^2}}.
\end{align}
Using $G_0(x,y)=e^{z_1(x-1)+z_2(y-1)}$ for randomply-coupled duplex ER networks, Eqs.~(3) and (4) are reduced to a single equation for $x_1=x_2=1-S$, so that the outbreak size $S$ is given by the solution of  
\begin{align}
1-S=e^{-(z_1 \tilde{T}_1 +z_2 \tilde{T}_2)S} .
\end{align}
Similarly for the epidemic probability, Eqs.~(5) and (6) are reduced to two coupled equations for $y_i=1-P_i$, given by 
\begin{align}
1-P_{1}&=e^{-(z_1 T_{s}P_1+ z_2 T_{d}P_2)}, \nonumber\\
1-P_{2}&=e^{-(z_1 T_{d}P_1+ z_2 T_{s}P_2)}.
\end{align}

\begin{figure}
\includegraphics[width=15.cm]{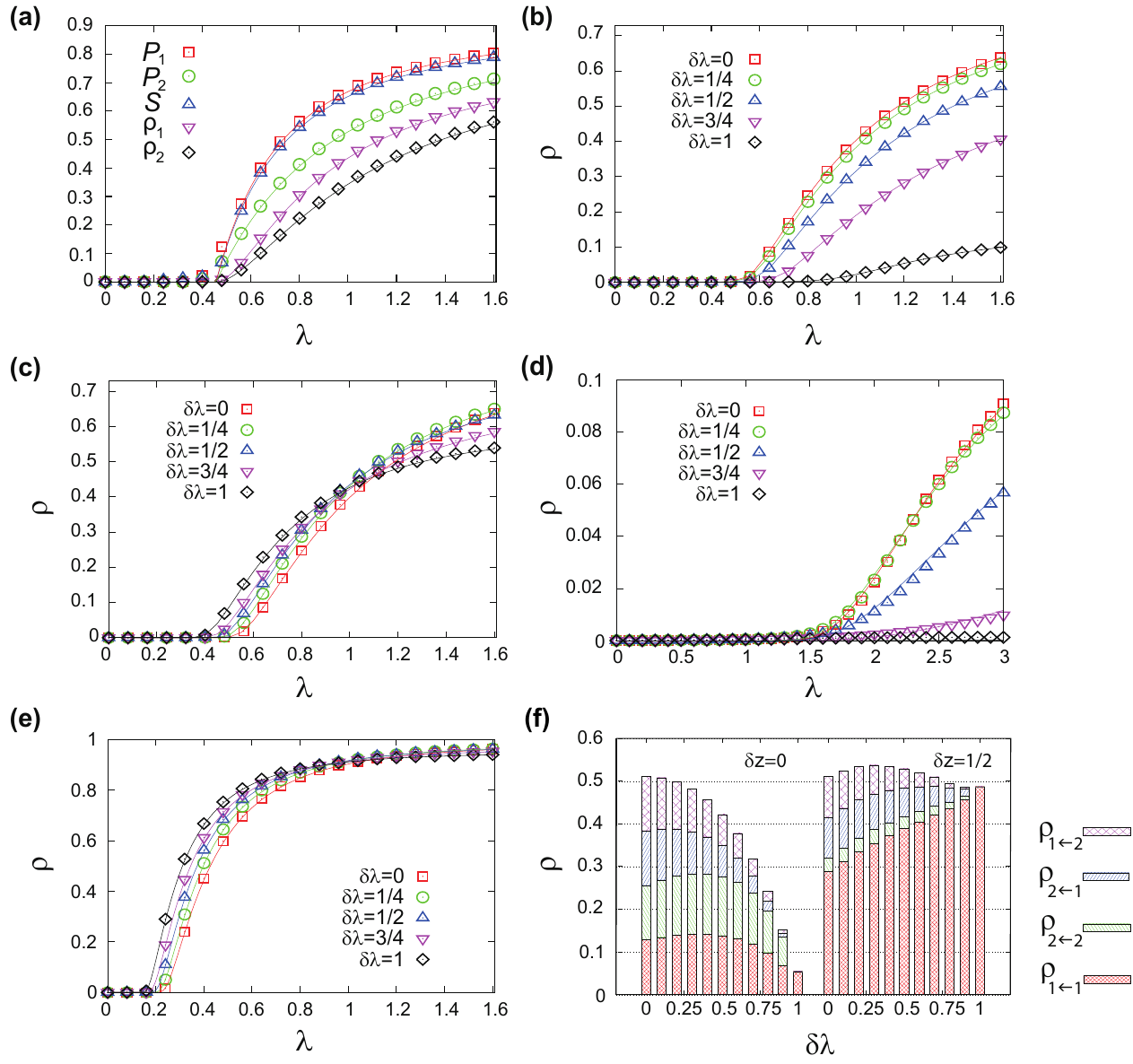}
\caption{
(a) Comparison of the analytical calculation (lines) and numerical simulation (symbols) results for the outbreak probabilities $P_1$ ($\Box$) and $P_2$ ($\circ$), 
the outbreak size $S$ ($\triangle$), and the prevalence
$\rho_1$ ($\triangledown$) and $\rho_2$ ($\diamond$), plotted 
as a function of $\lambda$. The results are for the duplex ER networks with $z_0=2.5$, $\delta z=1/2$, and $\delta \lambda=1/2$.
Numerical simulation results are obtained with $N=10^4$ nodes.
(b, c) The prevalence $\rho$ on duplex ER networks with $z_0 =2.5$, $\delta z=0$ (b) and $\delta z=1/2$ (c) for several values of cost parameter 
$\delta \lambda=0$ ($\Box$), $1/4$ ($\circ$), $1/2$ ($\triangle$), $3/4$ ($\triangledown$), and $1$ ($\diamond$).
Theoretical curves (lines) and numerical results obtained with $N=10^4$ nodes (points) are shown together.
(d, e) The plots of prevalence $\rho$ on duplex ER networks with $\delta z=1/2$ for $z_0=1.25$ (d) and $z_0=5.0$ (e) for several values of cost parameter 
$\delta \lambda$. Same symbols and lines as panels b and c are used. 
(f) The infection path profile in terms of the stacked histogram of the epidemic outbreak size different infection channels $\rho_{ji}$ for duplex ER networks of $z_0=2.5$ and $\delta z=0$ (left) and $\delta z=1/2$ (right), with $\lambda=1.2$ and various values of $\delta\lambda$.
}
\label{fig:er_fig3}
\end{figure}

As shown in Fig.~2a, for randomly-coupled duplex ER networks
the proposed analytical calculation results exhibit good agreement with the numerical simulation results even for the networks with moderate size $N=10^4$, supporting the validity of the analytic framework. Deviations observed near the epidemic threshold are due to the finite size.
Moreover the results in Fig.~2a manifest clearly that the outbreak probability $P_i$ and size $S$ can be different each other above the epidemic threshold.
It can also be noted that $P_1$ and $P_2$ can be different each other so that the outbreak probability and prevalence above the epidemic threshold does depend on the layer from which the infection is initiated.

\subsection*{Assessing the effect of layer-switching cost}
To assess the effect of layer-switching cost in minimal way, we employ a new parametrization for the infection rates and the mean degrees as follows. First, the infection rates are parameterized by 
\begin{align} \lambda_{s}=(1+\delta \lambda)\lambda \quad \textrm{and}  \quad \lambda_{d}=(1-\delta \lambda)\lambda~. \end{align}
Here $\lambda$ is the average infection rate and $0\le\delta\lambda\le1$ accounts for the level of layer-switching cost, such that $\lambda_s\ge\lambda_d$. By this parametrization we consider the scenario in which one could modulate the difference in $\lambda_s$ and $\lambda_d$ with $\delta\lambda$ while the average infection rate is kept fixed by the total amount of resource for information spreading. 
Similarly, the mean degrees of the two layers are parametrized as $z_{1}=(1+\delta z)z_0/2$ 
and $z_{2}=(1-\delta z)z_0/2$.
Here $z_0$ is the total mean degree of the two layers and $0\le\delta z\le 1$ quantifies the disparity in the link density of the two layers. 
By using this parametrization we aim to assess the effect of layer-switching cost as the relative link density of two layers changes while the total number of links is kept fixed.

To have the first sense for the effect of layer-switching cost, we take a look at the information spreading dynamics on duplex ER networks initiated from a type-$1$ transmission for several values of $\delta \lambda$.
Two values of link density disparity $\delta z=0$ and $\delta z=1/2$ for the same total degree $z_0=2.5$ (equivalently, $z_1=1.875$ and $z_2=0.625$) are chosen for comparison. 
As shown in Fig.~\ref{fig:er_fig3}, the effect of layer-switching cost  is multifaceted, depending on the network (parametrized by $\delta z$ here) as well as which aspect of information spreading one is interested in. 
For $\delta z=0$, the effect is rather simple: the layer-switching cost tends to hinder the information spreading, in that the larger $\delta\lambda$ is, the larger is the epidemic threshold $\lambda_c$ as well as the smaller is the prevalence $\rho$ (Fig.~\ref{fig:er_fig3}b). 
For $\delta z=1/2$, however, the effect of layer-switching cost is more intricate (Fig.~\ref{fig:er_fig3}c). 
As $\delta\lambda$ increases from zero, the epidemic threshold becomes smaller, meaning that the epidemic outbreak is facilitated near the threshold. On the contrary, larger $\delta\lambda$ yields smaller value of prevalence $\rho$ when $\lambda$ is sufficiently larger than the epidemic threshold.
For large enough $\lambda$ the system is well percolated, so large value of layer-switching cost causing confinement of epidemic spreading within the initial layer hinders the effective use of entire available network and thus produces suppressive effect.
What happens for small $\lambda$ is instead that the confined spreading within the denser layer due to large layer-switching cost becomes advantageous by avoiding the trapping of spreading in the sparse layer below percolation threshold. 
In this way, the layer-switching cost can lead to apparently counteracting effect depending on the average infection rate $\lambda$. 
Numerical simulation and theoretical calculation for different total mean degree $z_0=1.25$ and $z_0=5.0$ with $\delta z=1/2$, for which both the layers are unpercolating (percolating) for the former (latter),  show qualitatively similar pattern (Figs.~2d,e).

\subsection*{Infection channel profile}
The unequal usage of different transmission channels arises from the link density and the layer-switching cost. To quantify this we make the infection channel profile consisting of epidemic outbreak sizes due to each channel $\rho_{i\leftarrow j}$, which can be computed by the fraction of each transmission channels $T_{ij}$ is used during the information spreading process.
In Fig.~2f we show the infection channel profiles for the previously-examined two cases of duplex ER networks of $z_0=2.5$ with $\delta z=0$ and $\delta z=1/2$ (corresponding to Figs.~2b and c), respectively. We take $\lambda=1.2$, well above the threshold.   
For $\delta z=0$, as the layer-switching cost parameter $\delta\lambda$ increases the use of cross-layer transmission channels is suppressed more significantly whereas the intra-layer channels remain used in a similar level as long as $\delta\lambda<1$, illustrating clearly the simple detrimental role of layer-switching cost for $\delta z=0$.
For $\delta z=1/2$, on the other hand, 
the total epidemic size $\rho$ is more or less insensitive to $\delta\lambda$ 
while the composition of $\rho_{i\leftarrow j}$ significantly and systematically varies, with the intra-layer channel through denser layer $\rho_{1\leftarrow1}$ increasingly dominating the spreading as $\delta\lambda$ increases.

\begin{figure}
\includegraphics[width=14.5cm]{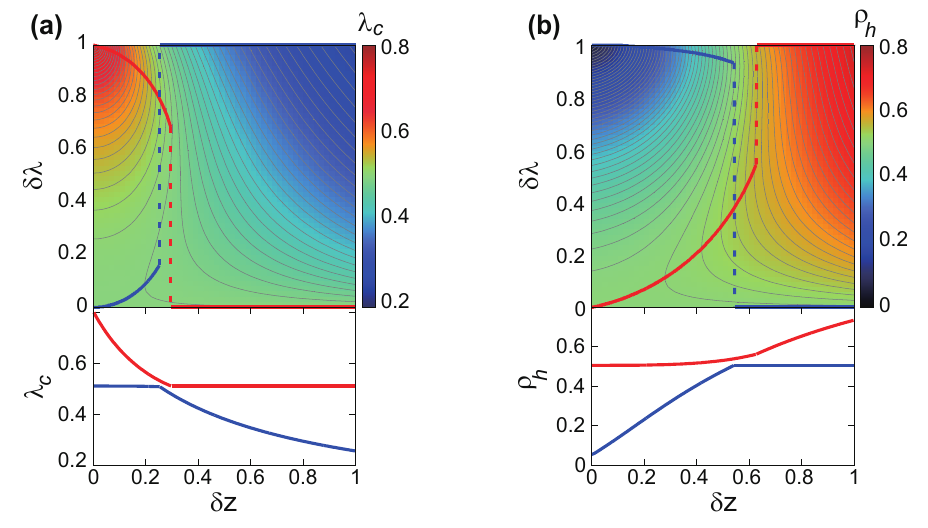}
\caption{
The epidemic threshold $\lambda_c$ (a) and the prevalence for $\lambda=1.2$ denoted as $\rho_h$ (b) as a function of $\delta z$ and $\delta\lambda$ for the randomly-coupled duplex ER network with $z_0=2.5$ are density-plotted in the top panels.
Thick red (blue) line in top panels is the trace of the loci of maximum (minimum) value of $\lambda_c$ (a) and $\rho_h$ (b) with respect to $\delta\lambda$ at the given $\delta z$, which can undergo discontinuous jump upon changing $\delta z$, indicated by the dashed line. 
At the bottom panels, the values of the maximal (red) and minimal (blue) $\lambda_c$ (a) and $\rho_h$ (b) with respective to $\delta\lambda$ are plotted as a function of the given $\delta z$.
}
\label{fig:er_outbreak_S_3d}
\end{figure}

\subsection*{Epidemic threshold and prevalence}
To establish a more comprehensive picture, we compute the epidemic threshold and  the prevalence for the full range of $\delta\lambda$ and $\delta z$. We specifically consider the prevalence $\rho_h$ computed for $\lambda=1.2$, well above the threshold, to address the situation where the level of available infection capacity is high enough for large-scale spreading. Therefore the two quantities could be a relevant measure for the efficacy of information spreading when the available infection capacity is tightly limited and sufficiently rich, respectively.

Plots of the epidemic threshold $\lambda_c$ and the prevalence $\rho_h$ for the duplex ER networks with $z_0=2.5$ are shown in Fig.~3. 
In Fig.~3a, we also indicate in the upper panel the loci of $\delta\lambda$ producing the largest threshold ($\lambda_c^{\rm max}$, red) and smallest threshold ($\lambda_c^{\rm min}$, blue) for given $\delta z$. The corresponding maximal and minimal $\lambda_c$ as a function of $\delta z$ is shown in the lower panel. 
The loci of $\delta\lambda$ for maximal and minimal $\lambda_c$ jumps abruptly at $\delta z = 0.297$ and $\delta z = 0.253$, respectively, which is accompanied by the discontinuity of the fist derivative in the plots of $\lambda_c^{\rm max}$ and $\lambda_c^{\rm min}$ vs.\ $\delta z$.

Similarly, in Fig.~3b, we display in the upper panel the loci of $\delta\lambda$ producing the largest (red) and smallest (blue) $\rho_h$ for given $\delta z$. The corresponding largest and smallest prevalence $\rho_h$ as a function of $\delta z$ is shown in the lower panel. 
The loci of $\delta\lambda$ for maximal and minimal $\rho_h$ also undergo abrupt jump at $\delta z = 0.672$ and $\delta z = 0.543$, respectively, which is associated with the discontinuity of the first derivative in the plots of $\rho_h^{\rm max}$ and $\rho_h^{\rm min}$ vs.\ $\delta z$.

Taken together, the effect of layer-switching cost demonstrated by Fig.~3 can be summarized as follows.
First, its effect is rather simple either when the two layers have similar density ($\delta z\lesssim 0.2$) or when the network density disparity is high enough ($\delta z \gtrsim 0.7$). In such cases, its effect is largely monotonic: either suppressive or facilitative for information spreading, albeit the effect is reversed for small and large disparity. On the other hand, for intermediate range of disparity ($0.2 \lesssim \delta z \lesssim 0.7$), the effect of layer-switching cost is no longer simple. Its effect is non-monotonous as well as it can accompany an abrupt, substantial discontinuity in the optimal parameter under a slight change of network density disparity. 
In sum, since the layer-switching cost tends not just to hinder cross-layer transmissions but also to promote 
intra-layer transmissions, the relative contribution and tradeoff 
between the two effects subject to the given network parameters and the level of available infection capacity can result in non-trivial and non-monotonic consequences to information spreading dynamics on multiplex networks, leading us to the concept of optimality \cite{yy}.

\begin{figure}
\includegraphics[width=12.5cm]{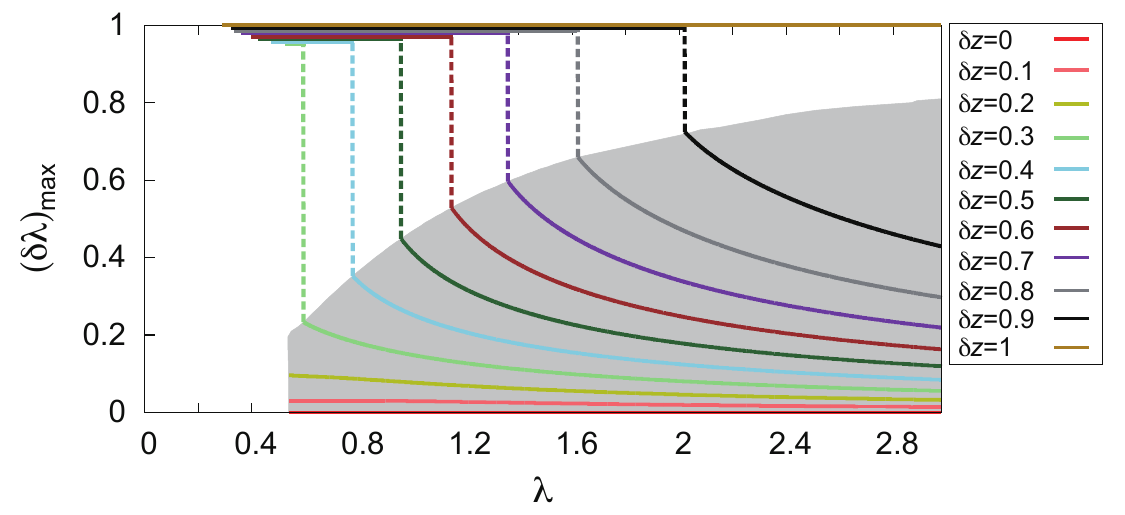}
\caption{
Plot of optimal cost parameter $(\delta \lambda)_{\rm max}$ for maximizing prevalence $\rho$ as a function of $\lambda$ with several values of $\delta z$.
The optimal parameter for information spreading with a limited resource (given $\lambda$) exhibits an abrupt discontinuous change for a wide range of $\delta z$.
Note that all the $(\delta\lambda)_{\rm max}$'s below the discontinuous transition point are exactly $1$ and overlapping; here we have purposefully split them next to each other for visual convenience.
}
\label{fig:optimal}
\end{figure}

\subsection*{Optimal layer-switching cost for maximal spreading}
The intricate effect of layer-switching cost allows formulation of many different optimization problems, contingent upon the objective of optimization as well as the given network parameters. In this section we analyze one particular optimization problem of finding the optimal layer-switching cost $(\delta\lambda)_{\rm max}$ that maximizes the prevalence for given total infection capacity dictated by $\lambda$, as an illustrative example. 

In Fig.~4, we show the optimal cost $(\delta\lambda)_{\rm max}$ as a function of $\lambda$ computed for duplex ER networks of $z_0=2.5$. 
When $\delta z < 0.280$, small value of $\delta\lambda$, that is, layer-indiscriminate spreading, is advantageous for any $\lambda$ above the threshold. 
On the other hand, when $\delta z> 0.280$, the optimal parameter $(\delta\lambda)_{\rm max}$ changes sensitively to $\lambda$. Moreover, it undergoes an abrupt discontinuous change  at some $\lambda$, whose location depends on $\delta z$, below which it is always advantageous to concentrate the spreading through the denser layer, that is, $(\delta\lambda)_{\rm max}=1$, down to the threshold $\lambda_c$ for spreading. The locations of the abrupt change in $(\delta\lambda)_{\rm max}$ constitute the boundary of shaded region in Fig.~4.
This example demonstrates explicitly how the non-analytical and discontinuous response of spreading dynamics to the layer-switching cost in multiplex networks can manifest generically in optimizing information spreading on multiplex networks. Optimization problems with other objectives such as minimizing the epidemic threshold for given network disparity can also be analyzed readily in this framework.

\begin{figure}
\includegraphics[width=14.5cm]{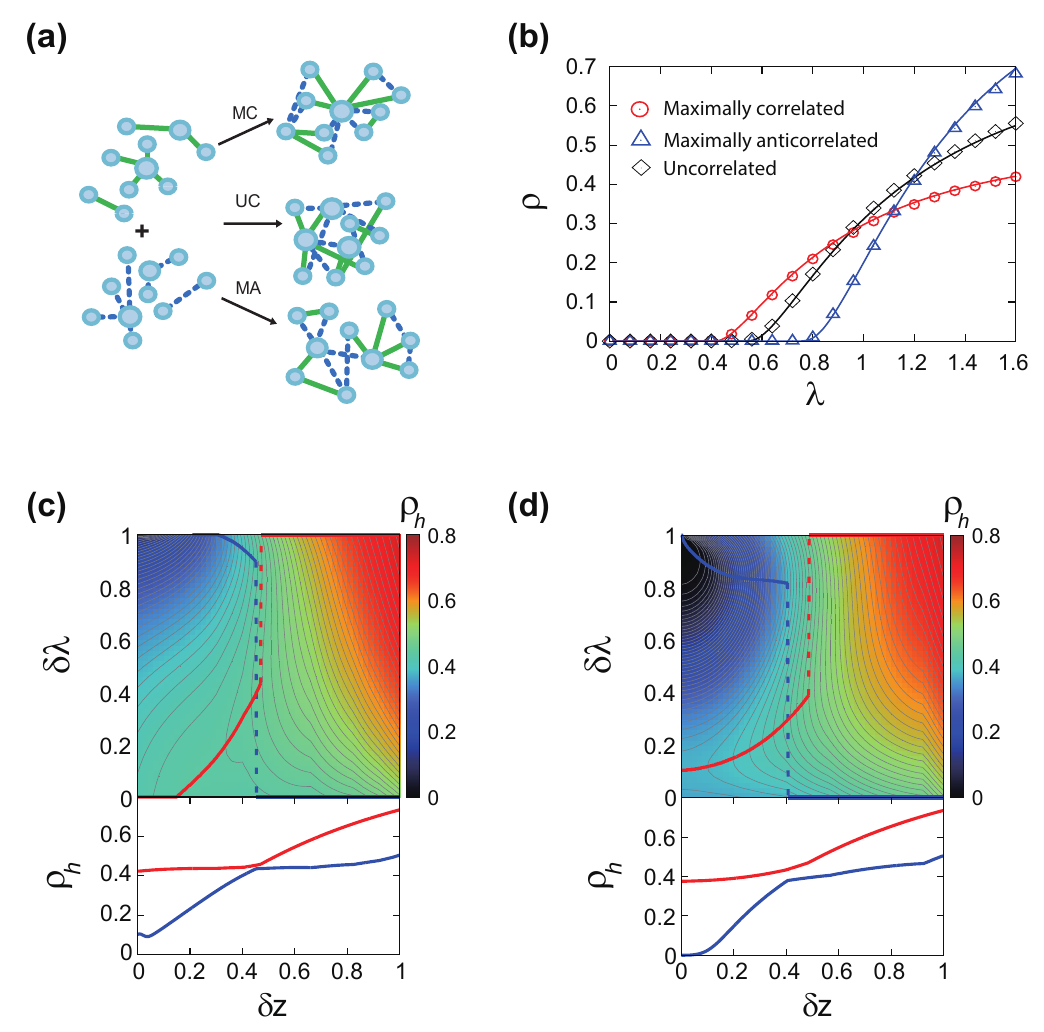}
\caption{
(a) Schematic cartoon illustrating the interlayer degree correlated couplings considered in the text.
MC stands for maximally correlated, UC for uncorrelaed, and MA for maximally anti correlated.
(b) Plots of epidemic prevalence $\rho$ on maximally-correlated ($\circ$), 
maximally-anticorrelated ($\triangle$), and uncorrelated ($\diamond$) duplex ER networks
with $z_0=2.5$ and $\delta \lambda=1/2$, as a function of $\lambda$ .
Numerical results obtained with $N=10^4$ nodes (points) and theoretical curves (lines) are in good agreement.
(c, d) The prevalence $\rho_h$ for $\lambda=1.2$ on the maximally-correlated (c) and maximally-anticorrelated (d) duplex ER networks with $z_0=2.5$, together with the maximum (red) and minimum (blue) $\rho_h$ plotted in the lower panels as a function of $\delta z$.
}
\label{fig:cor_er}
\end{figure}

\subsection*{Effect of interlayer degree correlations}
In investigating the effect of layer-switching cost  so far we have only considered the spreading processes on 
randomly-coupled multiplex networks, in which the degrees of a node in different layers are uncorrelated.
For many real-world networks, however, layers of a multiplex network often do not combine randomly. 
One of the simplest manifestation of the correlated coupling of multiplex layers is the interlayer degree correlation, that the degrees of a node at different layers are correlated, the effect of which has been examined for the robustness and controllability of multiplex networks~\cite{tan-cascading,min-robustness,control-correlated}.

We illustrate the effect of interlayer degree correlation by using duplex ER networks with
three representative cases of interlayer correlated coupling \cite{quark,min-robustness}:
Given two layers, we couple the layers in the maximally-correlated way by coupling the two nodes from each layer in their degree order; in the maximally-anticorrelated way by coupling the nodes in the opposite degree order; or just randomly in uncorrelated way (Fig.~5a).
Consequently, a node that is the hub in one layer is also the hub in the other layer for the maximally-correlated case, but it has the smallest degree in the other layer for the maximally-negative case. 

We show the prevalence plot for duplex ER networks of layers with equal mean degree $5/4$ ($z_0=5/2, \delta z=0$) and with layer-switching cost $\delta\lambda=1/2$ in Fig.~5b and for the entire range of $\delta\lambda$ and $\delta z$ in Figs.~5c,d, as illustrative example. 
In this case, the largest eigenvalue of the Jacobian matrix has the simple expression as
$\Lambda=T_s \kappa+  T_d \mathcal{K}$, where $\kappa$ and $\mathcal{K}$ are the self- and cross-second moments, respectively, of the joint degree distribution defined in Eq.~(1).
This indicates that the epidemic threshold should decrease with the interlayer degree correlation, codified by $\mathcal{K}$, confirmed in Fig.~\ref{fig:cor_er}b that $\lambda_c$ is lowest for the maximally-correlated case and highest for the maximally-anticorrelated case.
For the large enough $\lambda$, by contrast, the prevalence $\rho$ becomes largest for maximally-anticorrelated case and smallest for maximally-correlated case.
Therefore, the interlayer degree correlation facilitates the emergence of epidemics (lowering $\lambda_c$) 
but at the same time hinders the large-scale epidemic for high transmissibilities (smaller $\rho$),
reminiscent of the effect of degree assortativity in single-layer networks \cite{assortative}.
For the intermediate case, the response of the epidemic prevalence with respect to interlayer degree correlation is more complicated and dependent on details, as exemplified in Figs.~5c,d for the maximally-correlated (c) and maximally-anticorrelated (d) cases, respectively.

\section*{Empirical Twitter network}

We simulate the model on the empirical multiplex network constructed from Twitter data in Ref.~\citen{rumor_higgs}. This network consists of two layers, the retweet layer and the reply layer (Fig.~6a).
Each node is a Twitter account and two nodes are connected in the retweet layer if one ``retweets'' the other's tweet message at least once and in the reply layer if one ``replies'' to the other's tweet at least once. 
Although these two layers do not represent different communication media or platforms but different modes of usage of the common medium, Twitter, considering them from a functional point of view they are relatively autonomous, in that the flow of information is likely confined within the given mode and only occasionally crosses to the different mode: One is more likely to retweet the information seen from other's retweet message than to reply it to someone else. It can therefore by addressed, at least schematically, by the model with layer-switching cost.
Moreover, this dataset is one of the rare multiplex social network dataset which is publicly-available yet sufficiently large-scale, thus suitable for the modeling study.
The network contains $N=456,631$ nodes and the mean degrees of the two layers are $z_{\rm retweet}=3.21$ and $z_{\rm reply}=1.92$, corresponding to $z_0=5.13$ and $\delta z=0.25$. 
The degree distribution of each layer is fitted to a power law with the exponent $\approx -2.3$ (Fig.~6b).

We show the simulation results of the prevalence $\rho$ as a function of $\lambda$ (Fig.~6c) and the infection channel profile with $\lambda=1.2$ (Fig.~6d) for different $\delta\lambda$. On this network, the layer-switching cost is found advantageous for information spreading in wide range of $\lambda>0$,
until $\lambda$ becomes large enough $(\lambda\gtrsim2.5)$ when no appreciable differences are observed for different $\delta\lambda$. Infections through the retweet layer predominate the epidemic process, as this layer is denser. 
Compared with the model network results on duplex ER networks (Fig.~2), two notable structural features of the Twitter network are worth to be highlighted. First, the broad degree distribution of the Twitter network brings the epidemic threshold close to zero \cite{satorras}, so that the effect of layer-switching cost on changing the epidemic threshold is not observable. Secondly, structural organization of the two layers are not completely independent; rather they are highly correlated, since people tend to reply to someone who she/he had retweeted. In effect, there is prevalence of link overlap \cite{min-overlap} across the two layers. In the current dataset, $73.2\%$ of reply links are overlapping with retweet links and the size of giant connected component of the two-layer network is dominated by the retweet layer. This feature severely constrains the effect of outbreak suppression for large $\lambda$. 
Overall, the effect of layer-switching cost on the empirical Twitter data network is moderate yet non-negligible. Notably, when $\lambda$ is not too large ($\lambda\lesssim 1$), it can induce change in the  prevalence by as large as $10$ to $50\%$ (Fig.~6c, inset). The interplay of other higher-order structural features present in real-world networks, such as clustering and community structure \cite{gleeson-clustering} and evolution-driven correlation \cite{evolution,jungyeol}, for information spreading dynamics remains to be investigated further.

\begin{figure}
\includegraphics[width=15.5cm]{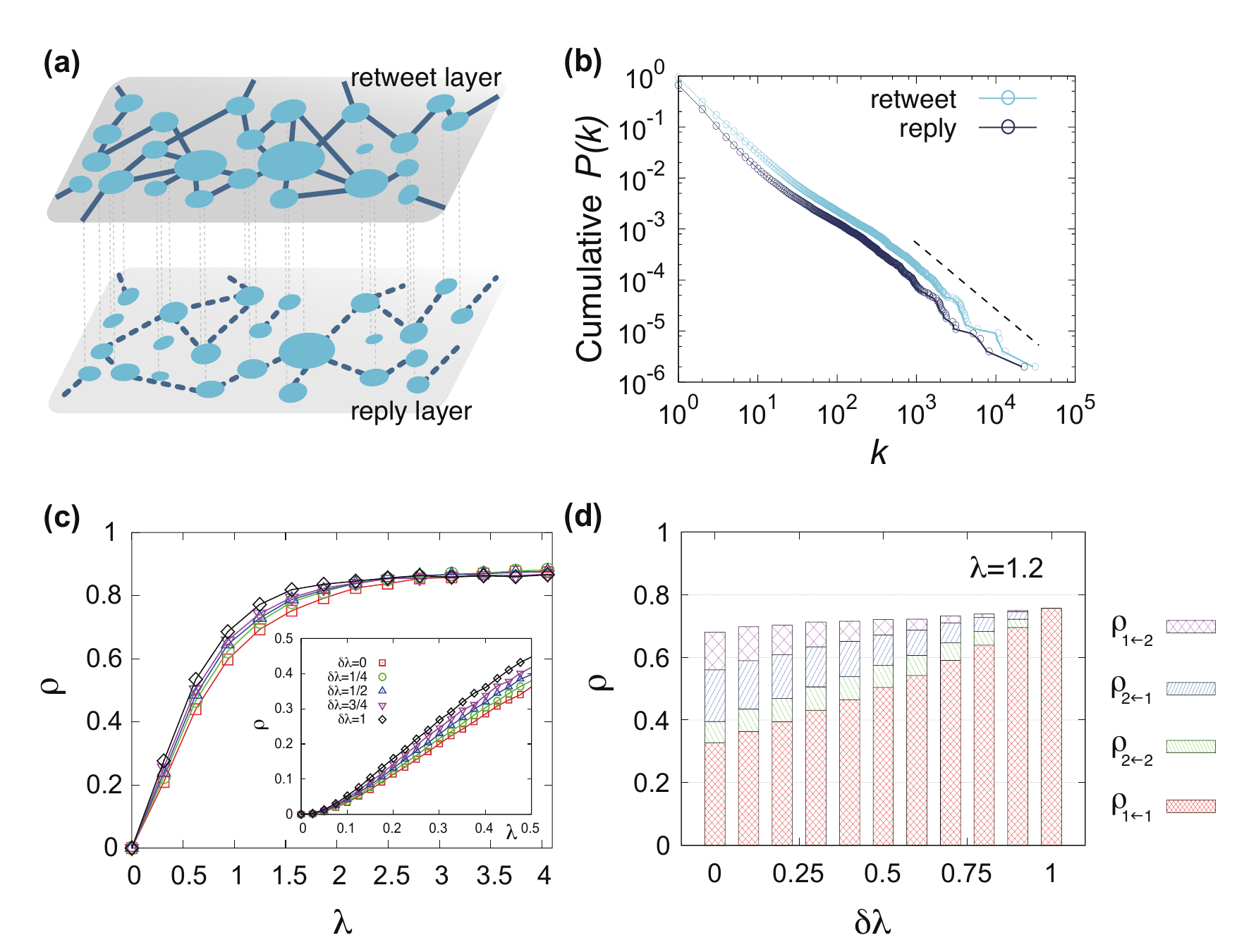}
\caption{
(a) Sample snapshot of a small portion of the 2-layer Twitter network. 
One can notice that the retweet layer is denser than the reply layer and a significant portion of links are overlapping across the two layers. Size of the node denotes the degree of the node in that layer.
The entire Twitter network data consists of $N=456,631$ nodes, with the mean degree of each layer being $z_{\rm retweet}=3.20$ and $z_{\rm reply}=1.92$, respectively.
(b) The cumulative degree distribution of the two layers in Twitter network. 
Data for both layers are fitted to a power law with an approximate exponent $\approx-2.3$ 
(as indicated by the dashed guideline with the slope $-1.3$). 
(c) Plots of the prevalence $\rho$ on the Twitter network for various values of $\delta\lambda$. 
(c, inset) A close-up of the prevalence plot for the range of small $\lambda$ ($0\le\lambda\le0.5$).
(d) The infection channel profile for the Twitter network with $\lambda=1.2$ and various values of $\delta\lambda$. 
}
\label{fig:empirical_histo}
\end{figure}

\section*{Infection Rates Dependent on Source Layer}

Although motivated originally to address the effect of layer-switching cost, 
the present model framework is applicable more broadly to the generic class of spreading processes involving layer-wise path-dependent trasmissibilities. 
One such case is where the infection rates are still path-dependent but determined primarily by the source layer. For example, one may have the infection rates parametrized as
\begin{align} \lambda_{j1}=(1+c)\lambda_{0} \quad \textrm{and} \quad \lambda_{j2}=(1-c)\lambda_{0} 
\end{align}
with $-1\le c\le 1$.
In the context of information spreading, such parameter setting may arise when a particular social layer (the layer 1 for $c>0$) has much higher credibility than the other so that the information received in that layer is taken more seriously and so more likely to be passed on through either layer whereas the information received in the other layer gets less attention and likely disregarded.  

We perform analysis of the model with infection rates given by Eq.~(13) on randomly-coupled duplex ER networks. 
The prevalence tends to be larger in this model than that for the original model with layer-switching cost, Eq.~(12), while the epidemic threshold tends lower (Fig.~7, to be compared with Fig.~3). 
The generalized application of our model framework put forward in this section also suggests a broader formulation of the optimization problem for information spreading in multiplex networks, beyond what has been discussed in comparison with numerical simulations part.
For example, the results obtained in this section show that the parameter setting for the source layer-dependent model, Eq.~(13), can be more effective in maximizing the information spreading than that with the layer-switching cost, Eq.~(12).

\begin{figure}
\includegraphics[width=14.5cm]{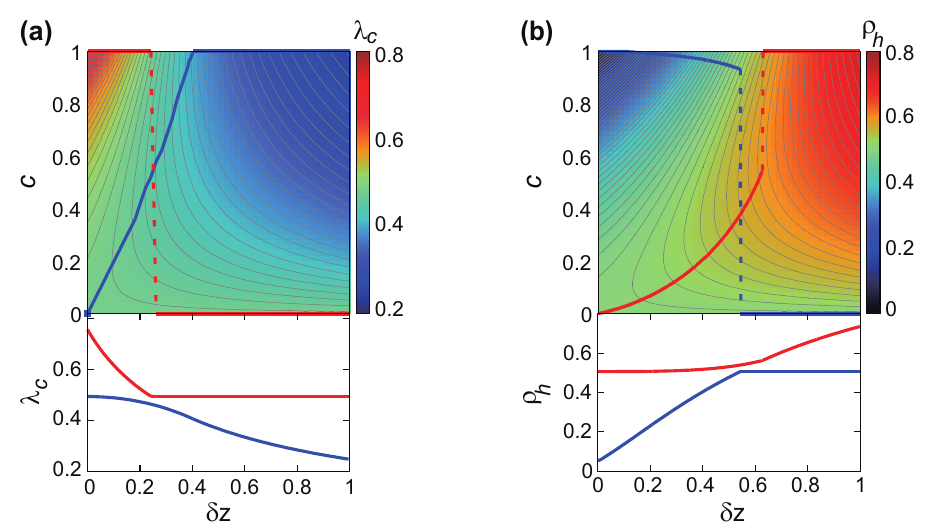}
\caption{
The epidemic threshold $\lambda_c$ (a) and the prevalence $\rho_h$ for $\lambda=1.2$ (b) as a function of $\delta z$ and $\delta\lambda$ for the source layer-dependent spreading model with infection rates Eq.~(13) on the duplex ER network with $z_0=2.5$ (top panels),
and the maximal and minimal values of corresponding observables with respective to $\delta\lambda$  as a function of the given $\delta z$ (bottom panels).
The red (blue) line in top panels represents the loci of maximum (minimum) of the corresponding observables at the given $\delta z$, same as in Fig.~3.
}
\label{fig:empirical_histo}
\end{figure}

\section*{Summary and Outlook}
In this paper, we have studied an information spreading model framework on multiplex networks with path-dependent transmissibility, paying particular attention to the effect of layer-switching cost.
We have formulated a generalized theory to deal with the path-dependent transmissibility
and illustrated how the epidemic threshold and prevalence could depend 
on the layer-switching cost, as well as on the network multiplexity factors such as the link densities of layers and the seed infection channel.
Optimal parameters for maximizing prevalence or minimizing epidemic threshold exhibit non-analytic behaviors, reminiscent of the abrupt structural transition in interconnected networks \cite{diffusion,structural}. 
Our formalism and results show that the seemingly benign factor of layer-switching cost is able to alter the macroscopic dynamic outcome in such nontrivial ways that the multiplex interactions cannot simply be reduced into a single aggregated layer \cite{moreno,KM_2014,diffusion}. 
According to our preliminary analysis, the effect of layer-switching cost is observed to be qualitatively similar for another classical epidemiological model, the SIS model \cite{anderson}, as well.
Therefore, the network multiplexity should explicitly be taken into account 
in order to understand and predict spreading dynamics accurately on multiplex networks. 
To a broader perspective, our results elucidate the impact of path-dependency in spreading process, which can arise also from the presence of memory in temporal networks, the effect of which has recently been studied \cite{diffusion1,diffusion2,naoki}. 
Finally, the multiplex information spreading model framework proposed in this paper furnishes us with a versatile platform for more realistic modeling of spreading processes involving layer-wise path-dependent trasmissibility on multiplex systems, offering a fertile ground for future study.

\section*{Acknowledgments}
This work was supported by the National Research Foundation of Korea (NRF) grants funded by the Korea government (MSIP) (No.\ 2011-0014191 and No.\ 2015R1A2A1A15052501).
K.-I.G. would also like to thank the APCTP for its hospitality during the completion of this work.

\section*{Author Contributions}
B.M. and K.-I.G.\ conceived the study. B.M., S.-H.G.\ and N.L.\ executed the research. B.M.\ and S.-H.G.\ performed the detailed analysis. B.M., S.-H.G.\ and K.-I.G.\ wrote the manuscript. 

\section*{Competing financial interests}
 The authors declare no competing financial interests.

\end{document}